\documentclass[twocolumn]{aastex631}

\usepackage{graphicx}	% Including figure files
\usepackage{amsmath}	% Advanced maths commands
\usepackage{amssymb}	% Extra maths symbols
\usepackage{comment}
\usepackage{braket}
\usepackage{booktabs}
\usepackage{tikz}

\newcommand\HI{H{\small{I}}}
\newcommand\Lyaf{Ly$\alpha$ forest}
\received{\today}
\revised{YYY}
\accepted{ZZZ}%\today}

\shorttitle{The BAO of the Ly$\alpha$ forest from simulations}
\shortauthors{Sinigaglia et al.}

\graphicspath{{./}{figures/}}

\begin{document}

%\title{The BAO of the Ly$\alpha$ forest from cosmological simulations: evidence for a negative shift of the BAO peak}

\title{The negative BAO shift in the  Ly$\alpha$ forest from cosmological simulations}

\correspondingauthor{Francesco Sinigaglia}
\email{francesco.sinigaglia@unige.ch}

\author[0000-0002-0639-8043]{Francesco Sinigaglia}
\affiliation{Département d’Astronomie, Université de Genève, Chemin Pegasi 51, CH-1290 Versoix, Switzerland}
\affiliation{Institut für Astrophysik, Universität Zürich, Winterthurerstrasse 190, CH-8057 Zürich, Switzerland}
\affiliation{Instituto de Astrof\'isica de Canarias, Calle via L\'actea s/n, E-38205, La  Laguna, Tenerife, Spain}
\affiliation{Departamento  de  Astrof\'isica, Universidad de La Laguna,  E-38206, La Laguna, Tenerife, Spain}

\author[0000-0002-9994-759X]{Francisco-Shu Kitaura}
\affiliation{Instituto de Astrof\'isica de Canarias, Calle via L\'actea s/n, E-38205, La  Laguna, Tenerife, Spain}
\affiliation{Departamento  de  Astrof\'isica, Universidad de La Laguna,  E-38206, La Laguna, Tenerife, Spain}

\author[0000-0001-7457-8487]{Kentaro Nagamine}
\affiliation{Theoretical Astrophysics, Department of Earth and Space Science, Graduate School of Science, Osaka University, \\
1-1 Machikaneyama, Toyonaka, Osaka 560-0043, Japan}
\affiliation{Theoretical Joint Research, Forefront Research Center, Osaka University, 1-1 Machikaneyama, Toyonaka, Osaka 560-0043, Japan}
\affiliation{Kavli-IPMU (WPI), University of Tokyo, 5-1-5 Kashiwanoha, Kashiwa, Chiba, 277-8583, Japan}
\affiliation{Department of Physics \& Astronomy, University of Nevada, Las Vegas, 4505 S. Maryland Pkwy, Las Vegas, NV 89154-4002, USA}

\author[0000-0002-5712-6865]{Yuri Oku}
\affiliation{Theoretical Astrophysics, Department of Earth and Space Science, Graduate School of Science, Osaka University, \\
1-1 Machikaneyama, Toyonaka, Osaka 560-0043, Japan}
\affiliation{Center for Cosmology and Computational Astrophysics, the Institute for Advanced Study in Physics,  Zhejiang University, China} 

%% Mark off the abstract in the ``abstract'' environment. 

\vspace{1cm}

\begin{abstract}
We present the first measurement of the \Lyaf{} BAO shift parameter from cosmological simulations. In particular, we generate a suite of $1000$ accurate effective field-level bias-based \Lyaf{} simulations of volume $V=(1 \, h^{-1} \, {\rm Gpc})^3$ at $z=2$, both in real and redshift space, calibrated upon two fixed-and-paired cosmological hydrodynamic simulations. To measure the BAO, we stack the three-dimensional power spectra of the $1000$ different realizations, compute the average, and use a model accounting for a proper smooth-peak component decomposition of the power spectrum, to fit it via an efficient Markov Chain Monte Carlo scheme estimating the covariance matrices directly from the simulations. We report the BAO shift parameters to be  $\alpha=0.9969^{+0.0014}_{-0.0014}$ and $\alpha=0.9905^{+0.0027}_{-0.0027}$ in real and redshift space, respectively. We also measure the bias $b_{\rm lya}$ and the BAO broadening parameter $\Sigma_{\rm nl}$, finding $b_{\rm lya}=-0.1786^{+0.0001}_{-0.0001}$ and $\Sigma_{\rm nl}=3.87^{+0.20}_{-0.20}$ in real space, and $b_{\rm lya}=-0.073^{+0.005}_{-0.004}$ and $\Sigma_{\rm nl}=6.55^{+0.23}_{-0.22}$ in redshift space. Moreover, we measure the linear Kaiser factor $\beta_{\rm lya}=1.39^{+0.24}_{-0.18}$ from the isotropic redshift space fit. Overall, we find evidence for a negative shift of the BAO peak at the $\sim 2.2\sigma$  and $\sim 3.5\sigma$ level in real and redshift space, respectively. %This result is qualitatively in agreement with measurements from other cosmological tracers in underdense regions, such as the cosmic void distribution. 
This work sets new important theoretical constraints on the \Lyaf{} BAO scale and offers a potential solution to the tension emerging from previous observational analysis, in light of ongoing and upcoming \Lyaf{} spectroscopic surveys, such as DESI, PFS, and WEAVE-QSO.
\end{abstract}


\section{Introduction} \label{sec:intro}

The Baryon Acoustic Oscillations (BAO) are standard rulers to measure cosmic distances 
\citep[][]{Sunyaev1970,Peebles1970,Seo2003,Blake2003}. They consist in the excess clustering observed at comoving $r\sim 150 \, h^{-1} \, {\rm Mpc}$ in the two-point correlation function and at different wavenumbers in the power spectrum of biased tracers of the dark matter field, and arise as a result of the propagation of acoustic density waves and the coupling between photons and plasma in the early Universe.

The first detection of the BAO independently by the \textit{Baryon Oscillation Spectroscopic Survey} (BOSS) and the \textit{2dF Galaxy Redshift Survey} (2dFGRS) collaborations using galaxy clustering \citep{Eisenstein2005,Cole2005} confirmed this prediction from the standard theory of cosmology. Since then, BAO studies have reported several other detections of the acoustic peak using different tracers of large-scale structure, supporting the paradigm described above 
\citep{Beutler2011,Anderson2012,Busca2013,Slozar2013,FontRibera2014b,Delubac2015,Kitaura2016b,Beutler2017,Bautista2017,Bourboux2020,Zhao2022}. The DESI \citep{Levi2013} and Euclid \citep{Amendola2018} surveys promise to achieve percent accuracy in the determination of the position of the BAO peak. Recently, DESI reported the detection of BAO over an effective volume of $V\sim 18 \, {\rm Gpc}^3$ and in six different redshift bins by relying on distinct tracers, with highest precision and significance of $\sim 0.83\%$ and $\sim 9.1\sigma$, respectively, at $z\sim 0.93$ \citep{DESI2024_KP4}.

The BAO peak position of the dark matter field in linear theory can be accurately predicted as a function of cosmology by means of a Boltzmann equation solver \citep[e.g.][]{Lewis2011,Blas2011}. However, the gravitational growth of structures induces the well-known distortions of the BAO. In particular, the nonlinear evolution of the density field broadens the BAO --- thus making the BAO measurements less precise --- and typically induces a systematic subpercent shift in the acoustic peak position whose magnitude and sign depend on the investigated cosmological tracer. To tackle the first problem --- which goes beyond the scope of this work --- a plethora of reconstruction techniques based on de-evolving cosmic structures in time have been put forward, and shown to successfully reduce the uncertainty on the BAO peak position without biasing the peak position itself \citep[e.g.][]{Eisenstein2007,Schmittfull2015}. As for the latter aspect, the evolved dark matter field itself, together with galaxies and haloes tracing high-density environments, have been reported to display a positive subpercent (typically of order $\lesssim 0.5\%$) BAO shift with respect to the prediction from linear theory, whereas the position of the BAO imprinted in low-density environments presents a negative shift of the same magnitude \citep{Sherwin2012,Achitouv2015,Kitaura2016b,Neyrinck2018,HernandezArguayo2020}. Finally, redshift space distortions (RSD) are also known to introduce a further positive contribution to the total shift in the BAO peak, also of order $\lesssim 0.5\%$ \citep[e.g.][]{Prada2016}.

In this context, the \Lyaf{} --- the ensemble of absorption lines imprinted on quasar spectra when light rays encounter intervening \HI{} clouds ---  has witnessed a recent growth in interest. In fact, the \Lyaf{} offers high-$z$ measurements of the BAO scale, and so a rare and robust probe of dark energy at high cosmic noon. Since the first measurements \citep{FontRibera2014b,Bautista2017,Bourboux2020} -- which reported a $4-5\%$ positive shift in the BAO peak along the parallel direction to the line of sight from the BOSS and eBOSS surveys --- the \Lyaf{} BAO scale has been in tension with the $\Lambda$CDM model. The most recent measurements of the \Lyaf{} BAO scale from the DESI survey \citep{DESI2024_KP6} sharpened the tension with eBOSS \Lyaf{} BAO results even more.

%BAO measurements of the \Lyaf{} from the BOSS and eBOSS surveys reported a $4-5\%$ positive shift in the BAO peak, thereby claiming a potential non-negligible tension with the $\Lambda$CDM model \citep{FontRibera2014b,Bautista2017,Bourboux2020}. 

On the other hand, theoretical studies of the \Lyaf{} BAO relying on simple bias models --- such as purely analytical studies \citep{Seljak2012,Chen2021} and approximate simulations \citep{White2010,Hadzhiyska2023} --- have not reported measurements of any potential shift of the BAO peak position.

In this work, we present the first determination of the \Lyaf{} BAO peak position and its shift with respect to the prediction from the linear theory of structure formation in $\Lambda$CDM models, based on accurate fast simulations, calibrated on full cosmological hydrodynamic simulations with baryonic physics of star formation and supernova feedback. 

The paper is structured as follows. In \S\ref{sec:sims} we present the two fixed-and-paired reference simulations that we use for calibration. In \S\ref{sec:approx_sims} we summarize the model that we adopt to generate the simulations used here. \S\ref{sec:bao_fit} outlines the procedure employed to fit the BAO from the set of simulations. \S\ref{sec:results} presents the results and develops a theoretical intuition of them. We conclude in \S\ref{sec:conclusions}.

% ***********************************
% ***********************************
% ***********************************

\section{Fixed-and-paired cosmological hydrodynamic simulations} \label{sec:sims}

In this work, we measure the \Lyaf{} BAO peak position from a set of fast \Lyaf{} simulations, generated using a bias model calibrated on two fixed-and-paired \citep{Angulo2016} cosmological hydrodynamical simulations. Such reference hydrodynamic simulations were run with the cosmological smoothed-particle hydrodynamics (SPH) code \texttt{GADGET3-OSAKA} \citep{Aoyama2018, Shimizu2019, Nagamine21}, a modified version of \texttt{GADGET-3} and a descendant of the popular $N$-body/SPH code \texttt{GADGET-2} \citep{Springel2005}. 
It embeds a comoving volume, $V=(500h^{-1}\text{Mpc})^3$, and $N=2\times1024^3$ particles of mass $m_{\rm DM}=8.43\times10^9h^{-1}\text{M}_\odot$ for DM particles and $m_{\rm gas}=1.57\times 10^9 h^{-1}\text{M}_\odot$ for gas particles.

The gravitational softening length was set to $\epsilon_g = 16 h^{-1}$\,kpc (comoving), but we allowed the baryonic smoothing length to become as small as $0.1\epsilon_g$. This means that the minimum baryonic smoothing at $z=2$ is about physical $533\,h^{-1}$\,pc. 
The star formation and supernova feedback were treated as described in \cite{Shimizu2019}.
The code also contains important refinements, such as the density-independent formulation of SPH and the time-step limiter \citep{Saitoh2009, Saitoh2013, Hopkins2013}. 

The main baryonic processes that shape the evolution of the gas are photo-heating, photo-ionization under the UV background radiation \citep{Haardt2012}, and radiative cooling. All of these processes are taken into account and solved by the \texttt{Grackle} library \citep{Smith2017}, which determines the chemistry for atomic (H, D and He) and molecular (H$_2$ and HD) species. The chemical enrichment from supernovae is also treated with the \texttt{CELib} chemical evolution library \citep{Saitoh2017}.
The initial conditions are generated at redshift $z=99$ using \texttt{MUSIC2} \citep{Hahn2021}, with cosmological parameters taken from the 2020 data release of the Planck collaboration \citep{Planck2018}. 

We use the output of the simulation at $z=2$ reading gas properties and compute the \Lyaf{} flux field by first obtaining the \HI{} optical depth $\tau$ by means of a line-of-sight integration as follows \citep{Nagamine21}: 
\begin{equation}\label{eq:tau}
    \tau(x_0) = \frac{\pi e^2}{m_e c} \int f \, \phi(x-x_0)\, n_{\rm HI}(x)\, dx \,
\end{equation}
where $e$, $m_e$, $c$, $n_{\rm HI}$, $f$, $x$ and $\phi$ denote, respectively, the electron charge, electron mass, speed of light in vacuum, \HI{} number density, absorption oscillator strength, line-of-sight coordinate, and the parametrization of the Voigt spectral line profile. Where necessary, relevant quantities (e.g., \HI{} number density) were previously interpolated on the mesh according to the SPH kernel of the simulation.  

We interpolated the \Lyaf{} fields in real and in redshift space onto a $1024^3$-cell cubic mesh --- i.e. at the native resolution of the simulation --- using a cloud-in-cell mass assignment scheme \citep{1981csup.book.....H}.

\section{Fast Lyman-$\alpha$ forest simulations} \label{sec:approx_sims}

We adopt the novel \textit{nonlocal Fluctuating Gunn-Peterson approximation} (NL FGPA) model \citep{Sinigaglia2023}, which builds on the standard FGPA. Briefly, the FGPA models the \Lyaf{} opacity $\tau$ as a power law of the underlying dark matter field overdensity $\delta$ as $\tau=A(1+\delta)^\gamma$ \citep{Bi1993,HuiGnedinZhang1997,Croft1998}. 
Starting from this point, the NL FGPA drops the assumption that a unique scaling relation holds over the full probed cosmological volume, and introduces instead a dependence on the cosmic web environment through the T-web classification \citep{Hahn2007}. After obtaining the gravitational potential $\phi(\vec{r})$ in Eulerian coordinates $\vec{r}$ by solving the Poisson equation $\phi(\vec{r})=\nabla^2\delta(\vec{r})$, the rank-2 gravitational tidal field tensor is defined as $\mathcal{T}_{ij}(\vec{r})=\partial_i\partial_j\phi(\vec{r})$, with eigenvalues $\lambda_1\ge\lambda_2\ge\lambda_3$. At this point, given the mesh discretization of the dark matter density field, a cell is classified as belonging to a \textit{knot} if $\lambda_1,\lambda_2,\lambda_3\ge0$, to a \textit{filament} if $\lambda_1,\lambda_2\ge 0, \lambda_3<0$, to a \textit{sheet} if $\lambda_1\ge0,\lambda_2,\lambda_3<0$, or to a \textit{void} if $\lambda_1,\lambda_2,\lambda_3<0$. The resulting nonlocal bias model for the \Lyaf{} becomes hence sensitive to the anisotropic clustering of the surrounding environment and encodes information up the third order of an Eulerian perturbative renormalized bias expansion \citep{McDonaldRoy2009,Kitaura2022}, i.e. all the orders which contribute to the clustering of the summary statistics studied herein \citep{WernerPorciani2020}. 

In addition to the parametrization of the geometry of the cosmic web, the NL FGPA includes two multiplicative threshold bias factors $\exp(\pm\delta/\delta^*)$ controlled by a scale parameter $\delta^*$ --- acting as a boosting (`$+$' sign) and cutoff terms (`$-$' sign) --- and a stochastic additive term $\epsilon$ sampled from a negative binomial distribution modeling a stochastic bias component. The final model reads:
\begin{equation}
\tau=A_i(1+\delta)^{\gamma_i} \, \exp(\delta/\delta^*_i) + \epsilon_i \quad , 
\end{equation}
%\begin{equation}
%\tau=A_i(1+\delta)^{\alpha_i} \, \exp(-\delta/\delta^*_{1,i})\, \exp(\delta/\delta^*_{2,i}) + \epsilon_i \quad , 
%\end{equation}
with $i=$\{knots, filaments, sheets, voids\}\footnote{Notice that we have reabsorbed the two free parameters $\delta^*_{1,i}$ and $\delta^*_{2,i}$ from \cite{Sinigaglia2023} in just one free parameter $\delta^*_{i}=1/\left(1/\delta^*_{2,i} - 1/\delta^*_{1,i}\right)$. Nonetheless, the degeneracy between $\delta^*_{1,i}$ and $\delta^*_{2,i}$ might be broken from additional observational/theoretical input in the future.}.

To include the effect of peculiar velocities and obtain the \Lyaf{} in redshift space (as is observed), we perform a cell-level mapping from real to redshift space \citep{Sinigaglia2021,Sinigaglia2022,Sinigaglia2023}, including a velocity bias contribution in the customary redshift space distortions formula \citep{Kaiser1984}:

\begin{equation}
    \vec{s} = \vec{r} \, + \, b_v \frac{(\vec{v}\cdot \hat{r})\, \hat{r}}{aH} 
\end{equation}
where $\vec{r}$ and $\vec{s}$ indicate real-space and redshift-space Eulerian coordinates, $b_v$ is the velocity bias, and $a$ and $H$ are the scale factor and the Hubble parameter at the studied redshift. The velocity field $\vec{v}=\vec{v}_{\rm  coh} \, + \, \vec{v}_{\rm qvm}$ is modeled as the sum of a large-scale coherent flow component $\vec{v}_{\rm  coh}$ --- which consists in the velocity field itself interpolated on the mesh --- and a small-scale quasi-virialized motion component $\vec{v}_{\rm qvm}$, a component randomly-sampled from a Gaussian distribution with zero mean and density-dependent standard deviation: $\vec{v}_{\rm qvm }\curvearrowright \mathcal{G}(0,\sigma)$, $\sigma=B(1+\delta)^\beta$, with $B$ and $\beta$ free parameters. The latter component is included because the mesh interpolation of the particle velocities smooths out the contribution below the scale of the cell, thereby damping the velocity dispersion component. Within the NL FGPA approximation, we always consider the dark matter velocity field instead of the one of the gas, even though we are modeling a baryon observable. This is done because the model needs to be generalized to any structure formation model, and approximated gravity solvers do not model any gas properties and yield just (approximated) dark matter velocities. The potential velocity bias between gas and dark matter arising from this assumption is then modeled and reabsorbed in the $b_v$ factor.

Eventually, the \Lyaf{} transmitted flux (i.e., normalized to the quasar continuum) is obtained as $F=\exp(-\tau)$.

We refer the reader to the paper presenting the NL FGPA model \citep{Sinigaglia2023}, for a thorough discussion of the improvements with respect to previous implementations of the FGPA.

Within this framework, we first calibrate the NL FGPA model on the reference simulations and reproduce their summary statistics. We obtain the dark matter field by applying the ALPT structure formation model \citep{Kitaura2013} with phase-space mapping \citep{Abel2012} on a grid of $N_p=256^3$ cells and a comoving volume $V=(500 \, h^{-1} \, {\rm Mpc})^3$ (physical cell resolution $l\sim 1.95 \, h^{-1} \, \rm{Mpc}$). In particular, we consider the same initial conditions as the reference simulations downsampled to the target resolution, and evolve them using ALPT down to $z=2$. After this first phase, to correct for the inaccuracy of the gravity solver and for the lack of resolution, we apply a rank-ordering nonlinear transform and force the approximated dark matter density field to follow the same distribution as the nonlinear evolved dark matter field from the N-body simulation. Afterward, we apply the NL FGPA model to predict the \Lyaf{}, both in real and in redshift space. We apply a rank-ordering transform also of the velocity field obtained with ALPT to the one from the cosmological simulation, in order to have a set of reliable nonlinear peculiar velocities to map the density field from real to redshift space. The parameters of the model are estimated by jointly fitting the flux probability distribution function (PDF) and the 3D power spectrum (for $k\le 0.5 \, h \, {\rm Mpc}^{-1}$) using a global optimizer implementing a genetic algorithm\footnote{\url{https://pypi.org/project/geneticalgorithm/}}. In particular, to prevent the risk of overfitting of just one cosmic realization, we jointly fit the summary statistics of the two fixed-and-paired simulations.  

\begin{figure}
    \centering
    \includegraphics[width=\columnwidth]{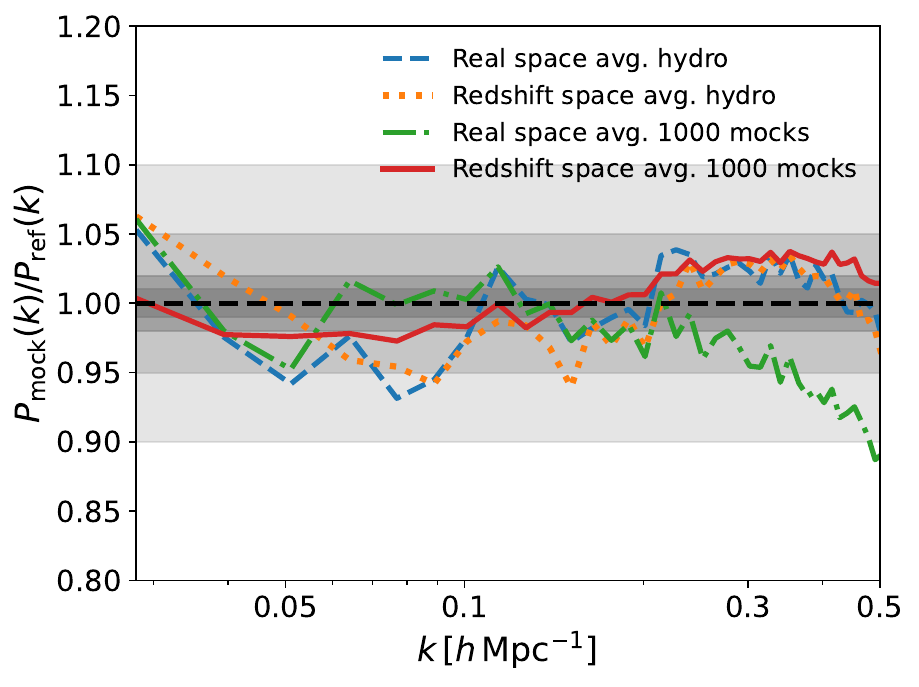}
    \caption{Ratios $P_{\rm mock}/P_{\rm ref}$ of three-dimensional \Lyaf{} power spectra. The blue dashed and orange dotted lines show the ratios of the average of the power spectra of the pair of simulations in real and redshift space, respectively. The green dashed-dotted and red solid lines display instead the power spectrum ratios of the mean of $1000$ mocks in real space and in redshift space, respectively. The gray dashed regions stand for $1\%$, $2\%$, $5\%$, and $10\%$ deviations, from the darkest to the lightest.}
    \label{fig:calibration}
\end{figure}

Figure~\ref{fig:calibration} shows the resulting calibration of the 3D power spectrum. The blue dashed and orange dotted lines show the ratios $P_{\rm mock}(k)/P_{\rm ref}(k)$ of the average of the power spectra  simulation pairs, in real and redshift space,  respectively. The average and maximum deviations in the power spectrum up to $k\sim0.5 \, h\, {\rm Mpc}^{-1}$ are $\sim 2.2\%$ and $\sim 5.3\%$ in real space, and $\sim 2.3\%$ and $\sim 6.2\%$ in redshift space. These figures are reduced to $\sim 1.8\%$ and $\sim 3.8\%$ in real space, and $\sim 2\%$ and $\sim 3.4\%$ in redshift space if we restrict the analysis to $0.1<k<0.5 \, h \, {\rm Mpc}^{-1}$, i.e. around the BAO scale. We also show power spectrum ratios for the mean of $1000$ mocks in real space (green dashed-dotted) and in redshift space (red solid), which achieves comparable accuracy to the one the calibrated power spectra in the $k$ range of interest. This demonstrates that the generalization of the calibration to random initial conditions yields substantially unbiased power spectra.

Using the calibrated parameters, we then generate $1000$ \Lyaf{} realizations of comoving volume $V=(1 \, h^{-1} \,  {\rm Gpc})^3$ (a factor $8$ larger than the one used for the calibration), both in real and in redshift space. In this case, the initial conditions are generated randomly, self-consistently convolving the white noise with the same input linear power spectrum as the simulations used for calibration.  

% ***********************************
% ***********************************
% ***********************************

\begin{figure*}
    \centering
    \vspace{-1cm}
    \includegraphics[width=\textwidth]{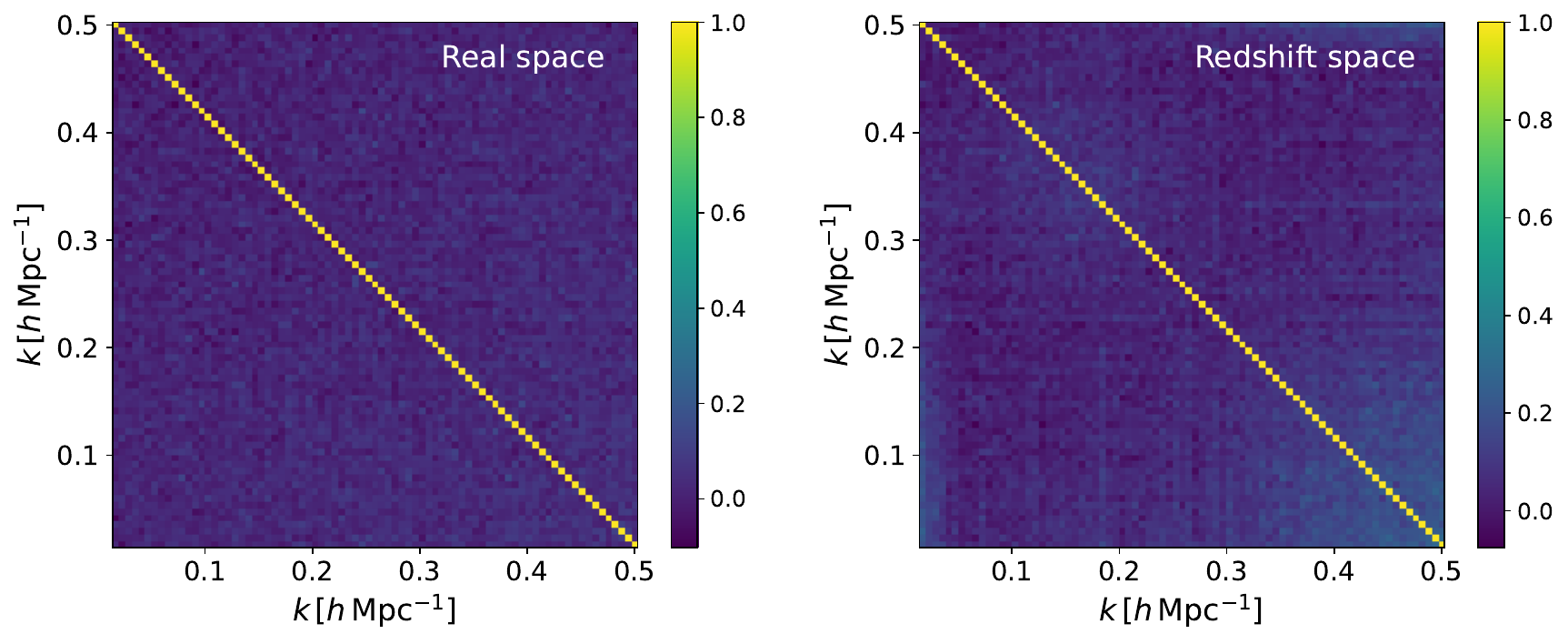}
    \caption{Correlation coefficient matrices in real (left) and redshift space (right), as obtained from the $1000$ fast \Lyaf{} simulations generated in this work.}
    \label{fig:covmat}
\end{figure*}

\section{Fitting the BAO of the Lyman-$\alpha$ forest} \label{sec:bao_fit}

We fit the \Lyaf{} three-dimensional power spectrum and derive the position of the acoustic peak and its shift parameter with respect to linear theory as follows.

We decompose the power spectrum into the sum of a smooth (no-wiggles) and peak (oscillating) components. In particular, we express the model for the power spectrum $P_{\rm model}(k)$ as \citep[e.g.][]{Bourboux2020}\footnote{We use the \texttt{Vega} package (\url{https://github.com/andreicuceu/vega}) to numerically compute the model.}:
\begin{equation}
    P_{\rm model}(\vec{k}) = b_{\rm lya}^2 \, (1+\beta_{\rm lya}\,\mu_k^2)^2 \, P_{\rm ql}(\vec{k}) \, F_{\rm nl}(\vec{k}) %\, G(\vec{k})
\end{equation}
where $\vec{k}=(k_\parallel, k_\perp)$ is the Fourier-space mode vector of modulus $k=|\vec{k}|$ and $\mu_k=k_\parallel/k$,  $b_{\rm lya}$ is the bias parameter, $\beta_{\rm lya}$ is the linear redshift space distortions parameter, $P_{\rm ql}$ is the quasi-linear matter power spectrum (described in detail below), $F_{\rm nl}$ is a transfer function accounting for the nonlinear growth of structures and modelling the nonlinear scales in the power spectrum. Differently from the main observational analyses carried out thus far in the literature, we do not include the correction term $G(\vec{k})$ accounting for the damping induced by the binning of the correlation function in $(r_\parallel, r_\perp)$ as we limit here our analysis to Fourier space and do not Fourier-transform the power spectrum into the correlation function in configuration space.

The quasi-linear power spectrum is modelled, as anticipated, as the sum of a smooth and a peak component:
\begin{equation}
    P_{\rm ql} = P_{\rm s}(k) + A(k) P_{\rm p}(k/\alpha)\exp\left(-\frac{k^2\Sigma_{\rm nl}^2}{2}\right)
\end{equation}
where $P_{\rm s}$ is the smooth component of the power spectrum obtained via empirical cubic spline interpolation, $P_{\rm p}=P_{\rm lin} - P_{\rm s}$ is the peak component with $P_{\rm lin}$ the linear power spectrum, $A(k)$ is a function describing the amplitude of the BAO, $\Sigma_{\rm nl}$ is a free parameter modeling the nonlinear broadening of the BAO peak, and $\alpha$ is the `shift' parameter, inducing a stretch ($\alpha<1$) or a compression ($\alpha>1$) of the power spectrum along the $k$ axis. Following \citet{Bourboux2020}, we reduce the function $A(k)$ to a $k$-independent parameter $A$, and find that this choice is sufficient to provide a good fit. Defining $\Delta^2(k)=k^3 P_l(k)/(2\pi^3)$ , the transfer function $F_{\rm nl}$ is based on the following model \citep{ArinyoPrats2015}:
\begin{equation}
    F_{\rm nl} = \exp\left\{ 
    %\left[ 
    q_1\Delta^2(k) 
    %+ q_2\Delta^4(k)\right] 
    \, \left[ 1- \left(\frac{k}{k_v}\right)^{a_v} \mu^{b_v}\right] - \left(\frac{k}{k_p}\right)^2\right\}
\end{equation}
with $q_1$, $a_v$, $k_v$, $b_v$, and $k_p$ free parameters\footnote{We have neglected the parameter $q_2$ used in \cite{ArinyoPrats2015}, whose impact is relevant only at $k$s manifestly larger than the ones probed in this work.}. Most of these parameters affect the shape of the \Lyaf{} power scale at highly nonlinear scales. 

We perform the fit as follows. First, we compute the covariance matrices in real and redshift space from the $1000$ simulations, which are shown in Fig.~\ref{fig:covmat}. The covariance matrix in real space (left) does not show any evident off-diagonal correlation, while the one in redshift space (right) features weak off-diagonal correlations at $k \gtrsim 0.4 \, h \, {\rm Mpc}^{-1}$. Afterwards, we stack the $1000$ \Lyaf{} realizations and fit the average power spectrum. We first determine the maximum likelihood best-fit solution, using the covariance matrices computed above. Then, we use the solution of the maximum likelihood estimation as an initial guess for a full Markov-Chain Monte Carlo sampling of the posterior distributions, assuming a Gaussian likelihood and flat priors around the maximum likelihood solution. The fit is performed over at $0.05<k<0.25 \, h \, {\rm Mpc}^{-1}$, as it is the $k$-interval which encompasses the three BAO peaks, as will be shown in Fig. \ref{fig:bao}. The further BAO peaks are found to be too degraded by nonlinear evolution to be reliably included in the fit.

\section{Results and discussion} \label{sec:results}

\begin{figure}
    \centering    \includegraphics[width=\columnwidth]{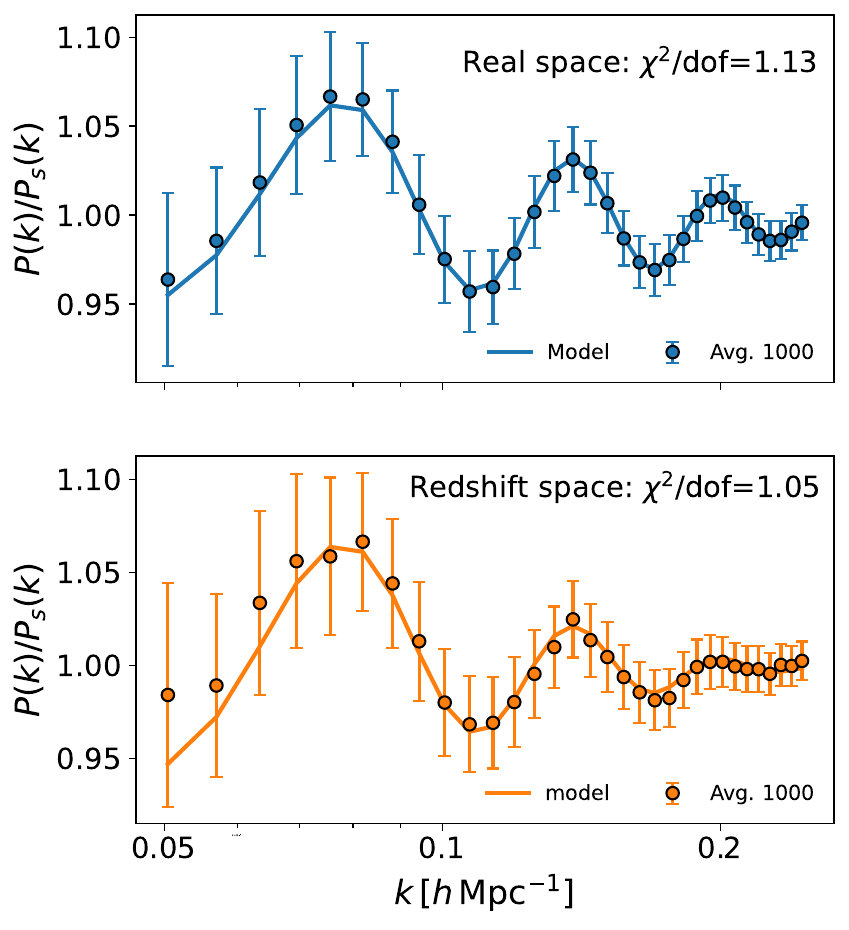}
    \caption{Results of the BAO fitting from the 3D \Lyaf{} power spectrum. This figure displays the ratio between the full \Lyaf{} power spectrum and the smooth component, corrected for the nonlinear transfer function $F_{\rm nl}$, in real (top) and redshift space (bottom). The data points are the average power spectrum from $1000$ simulations, the error bars indicate the uncertainties as computed from the diagonal element of the covariance matrix, and the solid line indicates the best-fit model. We also report the $\chi^2/$dof in the top right corner of each panel.}
    \label{fig:bao}
\end{figure}

We show the resulting fit of the BAO of the average \Lyaf{} power spectrum from our $1000$ realizations in Fig.~\ref{fig:bao}, in real (top) and redshift space (bottom). The fit yields a $\chi^2/{\rm dof=1.13}$ in real space and $\chi^2/{\rm dof=1.05}$ in redshift, i.e. the fits are regarded to be high quality. As can be noticed, in redshift space the third BAO peak is already damped by the nonlinear growth of structure and RSD. Nonetheless, we have explicitly verified that including it or not in the fitting procedure has a negligible impact on the results.

\begin{figure*}
    \centering
    \includegraphics[width=\textwidth]{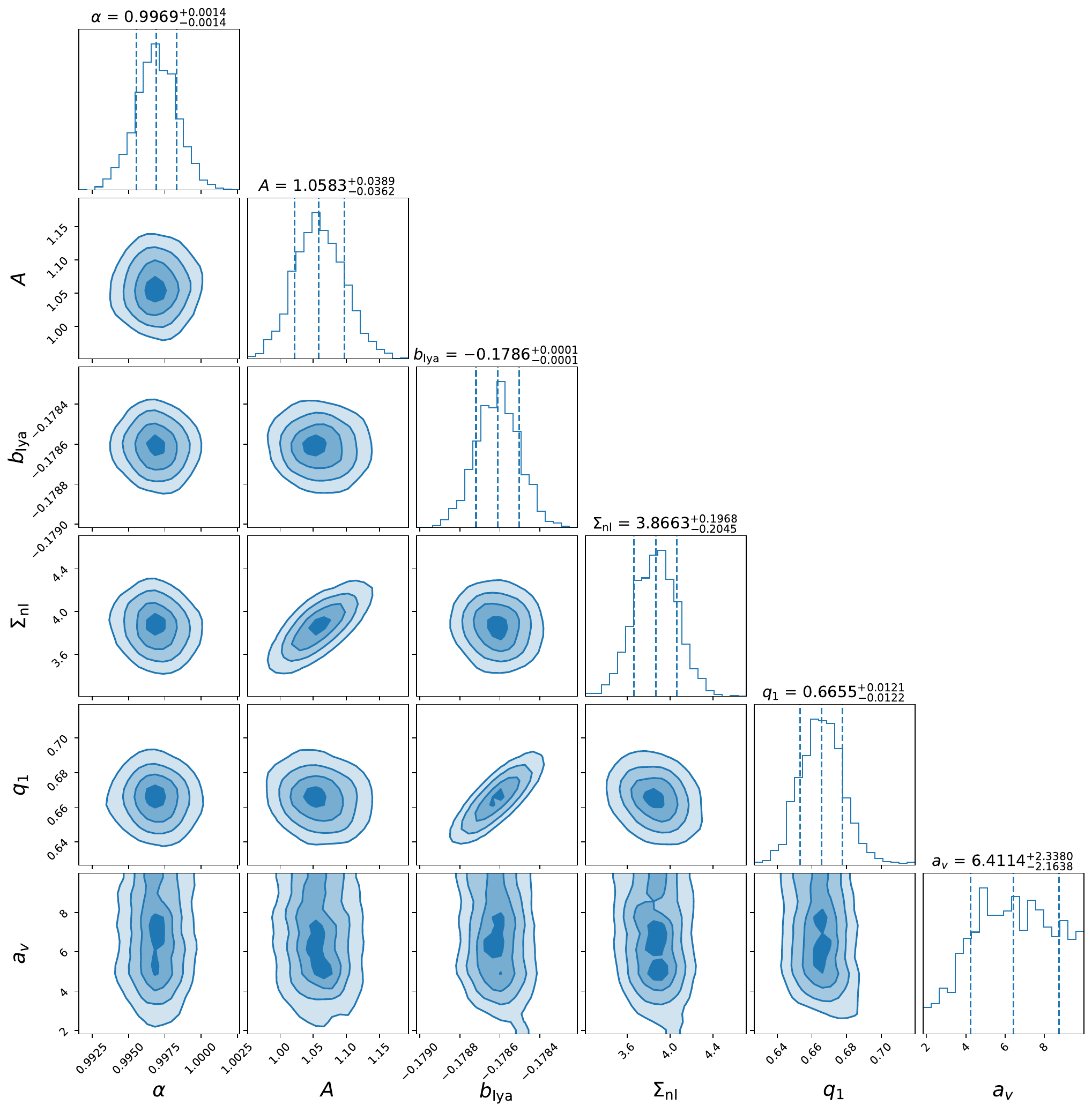}
    \caption{Posterior distributions of the model parameters in real space, obtained via Markov Chains Monte Carlo.}
    \label{fig:post_rspace}
\end{figure*}

\begin{figure*}
    \centering
    \includegraphics[width=\textwidth]{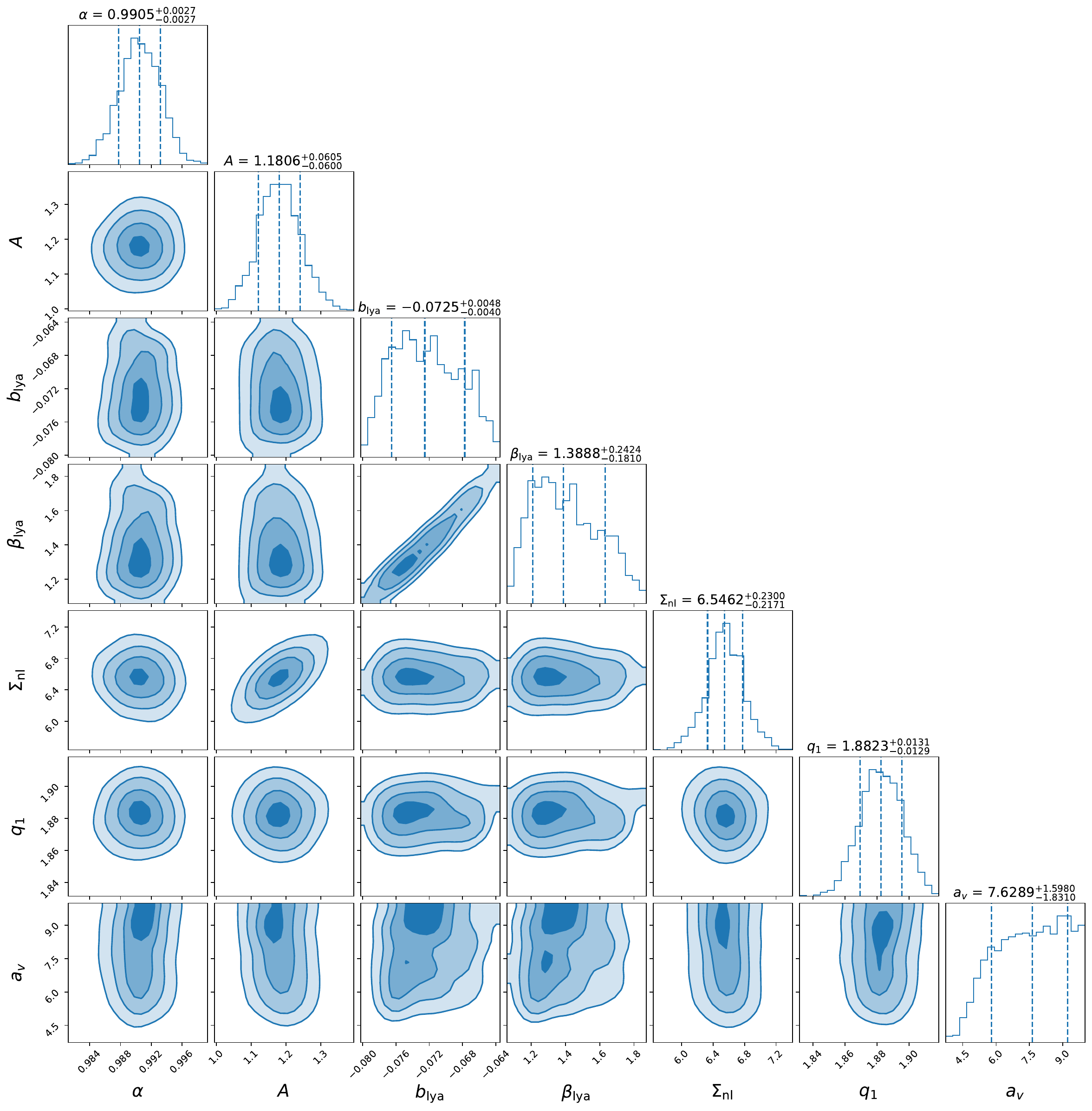}
    \caption{Posterior distributions of the model parameters in redshift space, obtained via Markov Chains Monte Carlo.}
    \label{fig:post_zspace}
\end{figure*}

Fig.~\ref{fig:post_rspace} and Fig.~\ref{fig:post_zspace} show the posterior distributions of the model parameters in real and redshift space, respectively. We quote as the best-fit value of each model parameter the $50${\it th} percentile of its posterior, and as associated uncertainties the combination of percentiles $84${\it th}--$50${\it th} (upper) and $50${\it th}--$16${\it th} (lower).  

We find the following BAO shift parameters:  $\alpha=0.9969^{+0.0014}_{-0.0014}$ in real space, $\alpha=0.9905^{+0.0027}_{-0.0027}$ in in redshift space. Overall, we find evidence for a negative shift of the BAO peak, both in real and in redshift space, at $\sim 2.2\sigma$ and  $\sim 3.5\sigma$, respectively. This result is qualitatively in agreement with measurements from other cosmological tracers residing in underdense regions, such as galaxies in cosmic voids, which have been shown to display a negative shift in the BAO peak. Moreover, we measure a difference at $\sim 2.2\sigma$ significance between the real and redshift space BAO shift, hinting towards an enhancement of the shift due to redshift space distortions, as previously known from the literature \citep[e.g.][]{Eisenstein2006,Prada2016}. These findings are compatible within statistical uncertainties with the most recent measurements of the \Lyaf{} BAO from the DESI collaboration \citep{DESI2024_KP6}. Remarkably, also DESI reported a negative median value of the \Lyaf{} BAO shift parameter, even though such a shift is not statistically significant given the measured uncertainties.  

We also measure explicitly for the first time the nonlinear broadening parameter $\Sigma_{\rm nl}$, finding $\Sigma_{\rm nl}=3.87^{+0.20}_{-0.20}$ in real space, and $\Sigma_{\rm nl}=6.56^{+0.23}_{-0.22}$ in redshift space --- in excellent agreement with the prediction at similar redshift from linear theory corrected for nonlinear evolution reported by \cite{Hadzhiyska2023} --- corresponding to $\sim 19.3\sigma$ and $\sim 30\sigma$ deviations from pure linear theory uncorrected for nonlinear evolution, respectively. Moreover, we find a difference between the real and the redshift space best-fit at $\sim 8.8\sigma$, with $\Sigma_{\rm nl}$ being larger in redshift space than in real space. This means that redshift space distortions contribute to the total BAO broadening with an additional BAO damping effect with respect to the effect given by nonlinear structure formation only, as expected. 

Furthermore, we report the measurement of the linear redshift space distortions parameter $\beta_{\rm lya}=1.39^{+0.24}_{-0.18}$ from the isotropic fit of the redshift-space 3D power spectrum.

To develop an analytical understanding of the findings obtained in this work, we rely on the Lagrangian displacement 
formalism presented in \cite{Eisenstein2006}. In particular, one can adopt the Zel'dovich approximation \citep{Zeldovich1970} to test the hypothesis that the BAO of the \Lyaf{} features a negative shift of the acoustic scale. 

Let us consider two positions in Lagrangian space $\vec{q}_1$ and $\vec{q}_2$ and compute their pairwise displacement $\vec{\Psi}_{12}$ projected along the initial separation vector $\vec{r}_{12}=\vec{q}_1-\vec{q}_2$:
\begin{equation}
    \Psi_{12,\parallel} = \frac{\vec{\Psi}_{12}\cdot\vec{r}_{12}}{r_{12}} = \int \frac{d\vec{k}}{(2\pi)^3}\, \delta_{\vec{k}} \, \frac{\vec{k}\cdot \vec{r}_{12}}{ik^2r_{12}} \, \left(e^{i\vec{k}\cdot\vec{q}_{1}} - e^{i\vec{k}\cdot\vec{q}_{2}}\right) \, .  
\end{equation}
This quantity tells what is the relative displacement of tracers under the action of gravity.

Let us now define and compute the mean displacement of cosmological tracers (e.g. galaxies) weighted by overdensity pairs:
\begin{equation}
\Psi^{p}_{\parallel} = \frac{\braket{\Psi_{12,\parallel }, \delta_{\rm{t},1} \, \delta_{\rm{t},2}} }{{\braket{\delta_{\rm{t},1} \, \delta_{\rm{t},2}}}} \, ,
\end{equation}
where $\delta_{\rm{t},i}$ is the Lagrangian tracers overdensity field at position $\vec{q}_i$, ${\braket{\delta_{\rm{t},1} \, \delta_{\rm{t},2}}}=\xi_{\rm t}(r_{12})$ is the correlation function. Notice that this quantity expresses the mean displacement weighted by the relative contribution to the correlation function. Also, we can express $\delta_{\rm t}$ as a  function of $\delta$ using a bias relation $\delta_{\rm t}=\mathcal{B}(\delta)$. Here, we assume a simple linear bias $\delta_{\rm t}=b\delta$ as we are interested in first-order effects, and we assume the bias to be just local, even though the bias could be arbitrarily nonlocal and the calculation would not change to leading order as nonlocal effects are higher-order effects \citep[e.g.][]{McDonaldRoy2009}. Eventually, as pointed out by \citet{Eisenstein2006}, the stochastic bias interpretation \citep{Dekel1999} does not change the results, as only $\delta_{\rm t}$ enters in the calculation to leading order when allowing the bias to be a stochastic distribution. 

Let us now consider the limit in which the correlation function is much smaller than the variance $|\xi(r_{12})|\ll \sigma^2$, where $\sigma^2=\braket{\delta^2}$, which holds true on sufficiently large scales, relevant for the large-scale linear bias framework we are working with. 

Under this assumption and performing all the calculations\footnote{We refer the reader to \cite{Eisenstein2006} for the details.}, one finds:
\begin{equation} \label{eq:final}
\Psi_\parallel^p= - \frac{2r_{12}J_3(r_{12})}{\sigma} \, \braket{\nu^2\nu_{\rm t}(\nu)} \, ,  
\end{equation}
where $J_3(r_{12})$ is the integral of the correlation function \citep{Peebles1980} and where we have defined the auxiliary variables $\nu=\delta/\sigma$ and $\nu_{\rm t}=\delta_{\rm t}/b\sigma$, i.e. we have rescaled the density fields by their standard deviation. 

Eq. \ref{eq:final} clearly shows that a positive linear bias $b>0$ induces a negative pairwise mean Zel'dovich displacement which corresponds --- as anticipated above --- to a positive shift of the acoustic scale when $\delta_{\rm t}>0$, as happens in the case of galaxies tracing dense environments. 

It is well-known that the linear bias $b_{\rm lya}$ of the \Lyaf{} must assume negative values, because the forest anti-correlates with the dark matter density field \citep[see e.g.][for an observational measurement of this value from eBOSS]{Bourboux2020}. One can also easily show this by means of the FGPA considering that $F=\exp(-\tau)$ and $\tau\sim(1+\delta)^\gamma$, which to linear order yields $F\propto -\delta$. For the \Lyaf{}, the condition $\delta_{\rm t}>0$ --- i.e. the regime which dominates the clustering --- is typically met in underdense or just mildly overdense regions, where the FGPA holds especially true, as the gas is not shock-heated \citep[e.g.][]{Lukic2015}.

Therefore, while computing the exact value for the shift goes beyond the scope of the paper and would inadequately neglect higher orders beyond linear, this simple argument can be used to provide a justification of the found negative shift ($\alpha-1$) parameter.

\section{Conclusions} \label{sec:conclusions}

In this work, we have presented the first quantitative study of the \Lyaf{} BAO shift from cosmological simulations. In particular, we generated $1000$ $V=(1 \, h^{-1} \, {\rm Gpc})^3$ fast accurate \Lyaf{} simulations at $z=2$. We first obtained the dark matter field using the ALPT structure formation model \citep{Kitaura2013}, and then applied the nonlocal FGPA prescription \citep{Sinigaglia2023} to paint the \Lyaf{} onto the dark matter fields. The free parameters of the nonlocal FGPA were calibrated in such a way to reproduce the 3D power spectrum of the \Lyaf{} from two fixed-and-paired full cosmological hydrodynamic simulations of volume $V=(500 \, h^{-1} \, {\rm Mpc})^3$.

We report $\alpha=0.9969^{+0.0014}_{-0.0014}$ in real space, and $\alpha=0.9905^{+0.0027}_{-0.0027}$ in redshift space, i.e., a negative shift of the BAO peak with respect to linear theory at a significance of 
$\sim 2.2\sigma$ and $\sim 3.5\sigma$, respectively. These findings are in agreement with the picture of an environmental dependence of the BAO shift emerging from the field of galaxy clustering. Moreover, we have shown that this result is also qualitatively consistent with perturbation theory predictions from a simple argument based on the Zel'dovich approximation. 

We conclude that this work sets new important theoretical constraints on the \Lyaf{} BAO scale and offers a potential solution to the tension emerging from previous observational analysis, in an exquisitely timely manner given the unprecedented precision achieved in the recent analysis from DESI Year 1 data \citep{DESI2024_KP6} and its relevance in the global cosmological analysis performed in combination with the data from galaxy clustering measurements \citep{DESI2024_KP7}.

\section*{acknowledgments}
The authors wish to warmly thank the reviewer for the valuable comments and suggestions he/she provided, which helped improve the quality of the manuscript.
The authors also thank Andrei Cuceu for his help in setting up the \texttt{Vega} model used in the fitting procedure and Alma Xochitl Gonzalez Morales for useful discussions in the early stage of this project. F.S. and F.S.K. acknowledge the Spanish Ministry of Economy and Competitiveness (MINECO) for financing the \texttt{Big Data of the Cosmic Web} project: PID2020-120612GB-I00/AEI/10.13039/501100011033 under which this work has been conceived and carried out, and the IAC for support to the \texttt{Cosmology with LSS probes} project. F.S. acknowledges the support of the Swiss National Science Foundation (SNSF) 200021\_214990/1 grant. K.N. is grateful to Volker Springel for providing the original version of \texttt{GADGET-3}, on which the \texttt{GADGET3-Osaka} code is based. Our numerical simulations and analyses were carried out on the XC50 systems at the Center for Computational Astrophysics (CfCA) of the National Astronomical Observatory of Japan (NAOJ), and {\sc SQUID} at the Cybermedia Center, Osaka University as part of the HPCI system Research Project (hp230089, hp240141).  This work is supported in part by the JSPS KAKENHI Grant Number 20H00180, 22K21349, 24H00002, 24H00241 (K.N.), and 21J20930, 22KJ2072 (Y.O.). 
K.N. acknowledges the support from the Kavli IPMU, World Premier Research Center Initiative (WPI), UTIAS, the University of Tokyo.  

\bibliography{sample631}{}
\bibliographystyle{aasjournal}

\end{document}